# A Closer Look at a Gamma-Ray Burst

**Stefano Covino** (INAF/Brera Astronomical Observatory)

A study of gamma rays produced when stars collapse or collide reveals details of the explosion mechanism, particularly the role of magnetic fields.

Gamma-ray bursts are among the most intriguing astrophysical events. Although short-lived, these explosions are the most luminous objects in the universe. However, the detailed mechanisms driving these bursts are still partly unknown. On page 1822 of this issue, the observations reported by *Mundell et al.* (1) will allow us to better understand the physical processes that power these celestial sources. By measuring the polarization of the electromagnetic radiation emitted immediately after a burst, Mundell et al. can help us unravel the role of magnetic fields in controlling the outflows produced by these explosions.

These burst events were first detected in the 1960s as intense and brief pulses of high-energy photons. They are now observed at photon energies across the whole electromagnetic spectrum and have been located at cosmological distances (that is, more than 10 billion light-years from Earth). They are also possible sources of gravitational waves, ultrahigh-energy cosmic rays, and neutrinos.

The most successful description of these events involves a very energetic outflow from an inner engine, either a massive star undergoing core-collapse or the merger of two compact objects (see the figure). Inhomogeneities in the outflow generate a prompt high-energy emission of gamma-rays, whereas the later interaction of the outflow with the matter surrounding the progenitor object generates a fainter and softer long-lasting emission, called the afterglow. One of the hottest open topics is to understand what drives the outflow, its composition, and its dynamics. And this is where the observations performed by Mundell et al. make an important contribution.

Gamma-ray burst outflows are extreme events. Initially they are ultrarelativistic, that is, the flow must have a velocity greater than 99.99% of the speed of light. This allows high-energy photons to escape from the very compact region where they are generated. The hypothesis accepted by most researchers assumes that the outflow is initially hot, with the expansion driven by its internal energy (2–4). This is known as the matter-dominated scenario, and serves as the reference model for outflows. The most popular alternative scenario requires that the outflow is driven by electromagnetic energy, which is called the Poynting flux–dominated scenario (5–7). (The Poynting flux is the flux of energy carried by an electromagnetic field) These two families of models differ in many aspects, although from the observational standpoint, they produce remarkably similar results.

These two scenarios also assume essentially the same emission processes, mainly synchrotron radiation (that is, radiation produced by electrons accelerated to relativistic velocities and usually characterized by a high degree of intrinsic polarization), although the presence of other components, such as inverse Compton scattering (that is, photons scattered to higher energies by the interaction with energetic electrons), has often been invoked. However, one important difference arises—if large-scale magnetic fields are present, they introduce a preferential direction in the system, breaking the spherical symmetry that can be assumed for the first phases of the outflow evolution. Independently of the details of the model, a general prediction for internal energy–dominated outflows is that during the first stages of the event, the polarized flux from the source should be almost negligible. In fact, due to the extreme relativistic effects acting at these stages, the observer sees only a small portion of the emitting outflow, and only very small anisotropies are expected (8–11). On the contrary, electromagnetic energy–dominated outflows could present initially a rather high polarized flux, perhaps as high as 30 to 40% (12–14).

Polarimetric observations performed during the first few minutes of gamma-ray burst evolution are therefore an effective observational test. If a strong polarized flux is measured, the hypothesis of a Poynting flux–dominated outflow is favored. Otherwise, hot, chaotic, internal energy–dominated outflows are to be preferred. Indeed, polarimetric observations of the high-energy emission of the prompt phase were attempted in the past, although none of the results were convincing enough due to the enormous technical difficulties. It is well known that polarimetry allows us to extract all the information carried by the photons we receive, in addition to the simple intensity. However, performing polarimetric observations of astrophysical sources has never been an easy task.

For gamma-ray bursts, these observations must be executed soon after the detection of high-energy photons in order to study a region still spatially close to the inner engine. The only way to cope with the unpredictable time and space location of a gamma-ray bursts, and be able to rapidly begin the observations, is with robotic telescopes, such as the 2.0-m Liverpool Telescope used by Mundell et al. Thanks to the rapid localization of gamma-ray bursts detected by the Swift satellite (15), Mundell et al. could observe the gamma-ray burst GRB 060418 only about 3 minutes after the burst. Their remarkable result is that only a rather low 8% upper limit for the linearly polarized flux at optical wavelengths was derived. Moreover, these observations were roughly coincident in time with the onset of the afterglow, as shown by Molinari et al. *(16)*. The lack of evidence for strong polarization just after the afterglow onset could suggest that large-scale magnetic fields are not likely playing an important role in driving gamma-ray burst outflows. However, if the magnetic fields are carried by the outflow ejected from the central source, as is commonly hypothesized, the prediction of the polarization during the afterglow onset depends on poorly known details of the magnetic energy transfer from the outflow to the shocked medium around the burst *(6, 17)*.

More observations tracing the early time evolution of the polarized flux, and an adequate modeling of these first phases of evolution of the outflow interaction with the matter surrounding the progenitor, will definitively settle the question. In addition, the possibility of carrying out polarimetric observations at high energies, in particular in the x-ray region, is an exciting future possibility. It could open the way for measurements during the prompt event, much closer to the progenitor, in a region where the possible effects of large-scale magnetic fields should be stronger.

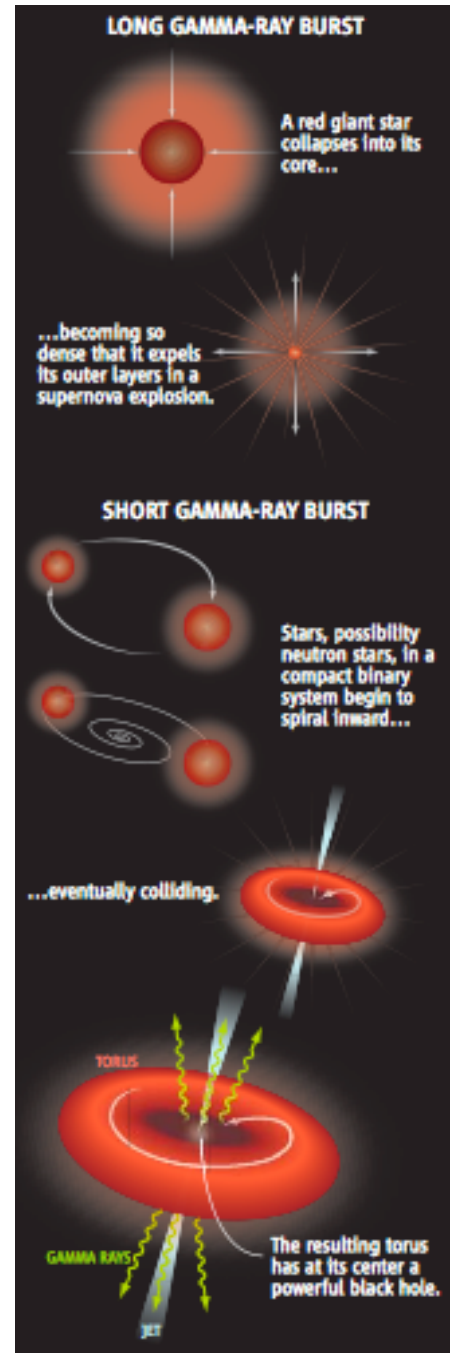

**Controlling the flow**. Gamma-ray bursts originate either from (top) collapsing stars or (middle) mergers of binary stars. The resulting high-energy event (bottom) creates ultrarelativistic outflows and very bright bursts of gamma-radiation.